# The redshift and geometrical aspect of photons

Hongjun Pan

Department of Chemistry

University of North Texas, Denton, Texas 76203, U. S. A.

**Abstract**

The cosmological redshift phenomenon can be described by the dark matter field fluid model, the results deduced from this model agree very well with the observations. The observed cosmological redshift of light depends on both the speed of the emitter and the distance between the emitter and the observer. If the emitter moves away from us, a redshift is observed. If the emitter moves towards us, whether a redshift, a blueshift or no shift is observed will depend on the speed vs. the distance. If the speed is in the range of $c(\exp[-\beta D] - 1) < v < 0$, a redshift is observed; if the speed equals $c(\exp[-\beta D] - 1)$, no shift is observed; if the speed v less than $c(\exp[-\beta D] - 1)$, a blueshift is observed. A redshift will be always observed in all directions for any celestial objects as long as their distance from us is large enough. Therefore, many more redshifts than blueshifts should be observed for galaxies and supernovae, etc in the sky. This conclusion agrees with current observations. The estimated value of the redshift constant $\beta$ of the dark matter field fluid is in the range of $10^{-3} \sim 10^{-5}$ /Mpc. A large redshift value from a distant celestial object may not necessarily indicate that it has a large receding speed. Based on the redshift effect of dark matter field fluid, it is concluded that at least in time average all photons have the same geometry (size and shape) in all inertial reference frames and do not have length contraction effect.

## 1. Introduction

In cosmology, the redshift is one of the most important observations to study the origin and the evolution of the universe and the motion of celestial bodies. Based on the redshift observation, in 1927, Belgian Priest Georges Lemaître was the first to propose that the universe began with the explosion of a primeval atom, such hypothesis is the origin of the Big Bang theory. The hypothesis was supported by Edwin Hubble (Hubble, 1929), he found that distant galaxies in every direction are going away from us with speeds proportional to their distance, which is based on the observation that the redshift is proportional to the distance. According to the Big Bang theory, as light from distant galaxies approaches the Earth there is an increase of space between the Earth and the galaxies due to the expansion of the universe, which leads to wavelengths being stretched, i.e., the light is expanding with the universe. The Big Bang theory becomes a dominant theory at current stage in study of the origin and the evolution of the universe. It can explain many important observations such as the redshifts, the ratio of light elements in the universe, cosmic microwave background radiation, etc. Tired light theory is an alternate explanation of the redshift effect. Tired light was first proposed in 1929 by Fritz Zwicky (Zwicky, 1929) who suggested that photons might slowly lose energy as they travel vast distances. The major problem associated with this theory is that there is no notable machnism causing such energy drop during its journey.

In 2004, the author proposed the dark matter field fluid model at the DPF2004 (Pan, 2005). In this model, the interstellar space is assumed to be, for simplicity, more or less uniformly filled with the dark matter field fluid which has fluid property and field property and all "baryonic" matter objects are saturated with such dark matter field fluid. Any motional celestial object will experience the dragging force of the dark matter field fluid. It is demonstrated that the current behavior and past evolution of Earth-Moon system can be described very well by this model (Pan, 2007) and the dragging effect of the dark matter field fluid dominates the evolution of the Earth-Moon system. This paper will extend the application of the dark matter field fluid model to the light traveling through space and compare the results with observations; the geometrical aspects of photons is also discussed based on this model.

## 2. The possible redshift effect of dark matter field fluid

In the proposed dark matter field fluid model (Pan, 2005 and 2007), a spherical body moving through the dark matter field fluid at low Reynolds number condition experiences the following dragging force F,

$$F = -6\pi\eta r^{1-n} mv \tag{1}$$

where η is the dark matter field fluid constant which is equivalent to regular fluid viscosity constant, r is the radius of the sphere, n is the constant rising from saturation effect, m is the mass of the sphere and the v is the moving velocity of the sphere. The direction of the force F is opposite to the direction of velocity. The equation of motion of the body is

$$m\frac{dv}{dt} = -6\pi\eta r^{1-n} mv \tag{2}$$

The equation (2) can be written as

$$\frac{d(mv)}{dt} = -6\pi\eta r^{1-n}(mv) \tag{3}$$

The momentum of the sphere is $p=mv$, so the equation (3) can be written as

$$\frac{dp}{dt} = -6\pi\eta r^{1-n} p \tag{4}$$

The equation (4) shows that under the dragging force of dark matter field fluid at low Reynolds number condition, the decrease rate of momentum of the spherical body is proportional to its momentum. It is further assumed that the general form of Eq. (4) can be applied to all ordinary matter objects (including photons) which move through the dark matter field fluid at low Reynolds number condition, i.e.,

$$\frac{dp}{dt} = -\alpha p \tag{5}$$

where α is a parameter depending on the geometrical characteristics of the object (such as the size, shape, etc), the dark matter field fluid constant. Eq. (5) is the law of motion for ordinary objects moving through the dark matter field fluid with low Reynolds number.

It is well known that a photon has momentum $p=h/\lambda$, where λ is the wavelength of photon and h is the Planck's constant. For a photon traveling through the dark matter field fluid, if the Eq. (5) is applicable, then,

$$h\frac{d\left(\frac{1}{\lambda}\right)}{dt} = -\alpha \frac{h}{\lambda} \tag{6}$$

and

$$\lambda = \lambda_0 e^{\alpha t} \tag{7}$$

where the $\lambda_0$ is the wavelength of the photon at time t=0, i.e., the wavelength of the photon which is just emitted by the emitter. The wavelength of the photon exponentially increases with the time it travels. This is the redshift effect of dark matter field fluid. The time t for the photons traveling from the emitter to the observer is

$$t = \frac{D}{c} \tag{8}$$

where the $D$ is the distance from the emitter to the observer and $c$ is the speed of light. Let $\beta = \alpha/c$, the redshift constant of dark matter field fluid, the Eq. (7) can be written as

$$\lambda = \lambda_0 e^{\beta D}. \tag{9}$$

By convention, the redshift z is defined as

$$z = \frac{\lambda - \lambda_0}{\lambda_0}. \tag{10}$$

So the redshift caused by dark matter field fluid is

$$z = e^{\beta D} - 1. \tag{11}$$

When the $\beta D \ll 1$, Eq, 11 reduces to the regular cosmological redshift formula

$$z = \beta D = \frac{\alpha}{c} D. \tag{12}$$

Eq. 12 indicates that for a sufficient short distance, the cosmological redshift is directly proportional to the distance. The conventional cosmological redshift formula is

$$z = \frac{H}{c} D \tag{13}$$

where the H is the Hubble constant. One can see that $\alpha$ is equivalent to H. Although the Eq. 12 and Eq. 13 are the same in form, their physical meanings are completely different. In the Eq. 12, the redshift is caused by the dragging effect of dark matter field fluid and the parameter $\alpha$ is a measure of the dragging effect of the dark matter field fluid on the light and has a unique value; but in the Eq. 13, the redshift is caused by universe expansion or the stretching of the space.

The emitter, however, may move with speed v relative to the observer, such motion has a Doppler effect on the emitted light (v $\ll$ c)

$$\lambda_0 = \lambda_r (1 + \frac{v}{c}) \tag{14}$$

where the $\lambda_r$ is the wavelength of the light when the emitter is at rest. Therefore, the actual wavelength of light detected by the observer is

$$\lambda = \lambda_r (1+\frac{v}{c})e^{\beta D}. \tag{15}$$

The observed redshift is

$$z = \frac{\lambda - \lambda_r}{\lambda_r} = (1+\frac{v}{c})e^{\beta D} - 1. \tag{16}$$

Obviously, the relativistic Doppler formula has to be used when the v is close to c. Eq. 16 indicates that the observed redshift z depends on not only the speed of emitter but also the distance. For sufficiently short distance, the Doppler motion effect may dominate the redshift z; for a large distance D, the contribution from the dragging force of the dark matter field fluid may dominate the redshift z.

Fiq. 1 shows how the redshift z varies with the speed of the emitter with parameter βD=0.05. By convention, if z is positive, it is redshift; if z is negative, it is blueshift. Referring to Fig. 1, the line intercepts with the speed axis v/c at (exp[-βD]-1), no at zero. When the emitter (such as galaxies, supernova, pulsars, etc) moves away from the observer (on the Earth), v is positive, both the dragging effect of dark matter field fluid and the effect of motion (Doppler effect) have positive contribution to z, z is positive, so a redshift is observed. When the speed of the emitter is zero, only the effect of dark matter field fluid contributes to z which equals to (exp[βD]-1), and a redshift is observed. When the speed is negative, i.e., the emitter moves toward the observer with the speed in the range between (exp[-βD]-1) and 0, the positive contribution by the dark matter field fluid is greater than the negative contribution by the motion, z is still positive, i.e., a redshift is observed, so the observer can not know which direction the emitter moves based on only the sign of z. When the speed of the emitter equals to (exp[-βD]-1), the positive contribution from the dragging effect of dark matter field fluid equals the negative contribution from the motion effect, the two effects cancel each other, no shift is observed. When the speed of the emitter is at the left side of the (exp[-βD]-1), the negative contribution from motion effect is greater than the positive contribution from the dark matter field fluid, a blueshift is observed. The observer knows that the emitter moves towards to him/her. When the distance D increases, the interception (exp[-βD]-1) moves to the left, covers a larger redshift range, a higher negative speed is needed in order to observe a blueshift. According to this result, due to the redshift effect of dark matter field fluid, much more redshifts than blueshifts should be observed for the celestial objects in the sky and a redshift (z > 0) is always observed for any celestial objects in all directions as long as their distance is sufficiently large. This conclusion is exactly what is observed. So far only few galaxies show blueshifts, the most famous being M31 galaxy (Andromeda). As M31 is our near neighbor and the distance is relatively short, the speed is at the left side of (exp[-βD]-1) on Fig. 1. In this case, the redshift associated with the dark matter field fluid is less than the blueshift from the motional Doppler effect. Most galaxies are much further away from us. It is possible that some of those far away galaxies may move towards us, but the redshift effect of dark matter field fluid is greater than the blueshift of the motional Doppler effect, the final observed shifts are to the red. The significance of the Eq. (16) is that a large value of the redshift z does not necessarily means that the distant emitters(galaxies, supernovae, etc) have a large receding speed. The speed deduced from a redshift z using conventional Doppler formula or conventional cosmological redshift formula (Hubble's law, v = HD) will be overestimated for v ≥ 0, misleading for c(exp[-βD]-1)≤ v ≤ 0, and underestimated for v < c(exp[-βD]-1). Note, the speed v here is the actual speed of the emitters relative to the observer at the moment it emits the light.

The parameter β will be one the important parameters of the nature. Finding the value of β is a challenge. The redshift z of a remote celestial object can be accurately measured; however, accurately measuring the speeds and distances of the distant celestial objects is difficult. We can roughly estimate the range of β. According to the available data, the M31 galaxy has a redshift z = -0.000991 (Huchra *et al*. 1999) and distance about 2.9 million light years which equals 0.889 Mpc (mega parsec). The speed of M31 would be -297 km/s if only based on the Doppler effect. This speed will be underestimated according to the Eq. 16. If the actual speed is -600 km/s (most likely exaggerated), then, β= 1.14 × $10^{-3}$ /Mpc; if the actual speed is -300 km/s, then, β = 1.09 × $10^{-5}$ /Mpc. So β is probably in the range of $10^{-3}$ to $10^{-5}$ /Mpc. According to Eq. 12 and Eq. 13, β = H/C = 2.5×$10^{-4}$ /Mpc with assumption of H = 75 km/s/Mpc, it is just in the middle of the estimated range. Therefore, the current value of Hubble's constant can be used as a good guess for the β. With this estimated value of β, the speed of M31 towards us is about -363 km/s. When the sufficiently accurate value of β and the distance of the objects are found, it will be possible to calculate the speed of the objects with the observed value of z.

Theoretically, the mechanism of the redshift effect by the dark matter field fluid model can be tested by observing the change of the redshift values of remote celestial objects over the time. As indicated above that when the emitter moves toward the observer with the speed in the range between (exp[-βD]-1) and 0, the positive contribution by the dark matter field fluid is greater than the negative contribution by the motion, z is still positive, i.e., a redshift is observed. However, the value of the redshift z will decrease with time as long as the emitter keeps moving towards the observer, and eventually it will become blueshift when the distance between the emitter and observer is short enough that redshift by dark matter field fluid is less than the blueshift by the motional Doppler effect. However, it will take very long time for any detectable change with reasonable accuracy. For example, it will take about 97 million years for the blueshift of M31 changing from -0.000991 to -0.001000 based on the current data and assuming $\beta = 2.5 \times 10^{-4}$/Mpc.

As indicated in the previous paper (Pan, 2005), the dark matter field fluid may have thermal property with temperature about 2.7 K, the observed cosmic microwave background radiation is the black body radiation of the dark matter field fluid at 2.7K. It is certain that the density distribution and the thermal temperature of the dark matter field fluid are not uniform through the space and not constant in all time scale. The uneven distribution of density and temperature of the dark matter field fluid in the space causes uneven black body radiation which could be the origin of the observed anisotropies of cosmic microwave background radiation. Furthermore, it is very possible that the dark matter field fluid may be converted to other type of matter by certain physical mechanisms, or vise verse, which causes the change of density and temperature distribution, therefore, the current distribution of the density and the temperature of the dark matter field fluid could be different from remote past.

It is interesting to notice that the redshift effect of dark matter field fluid is another version of "Tired light" model. But in this model, it is clear that the mechanism to cause the energy loss for photons to pass through the dark matter field fluid is the dragging force, and their wavelength becomes increasingly longer as they do so. No one ever thinks before that there is any relationship between the evolution of the Earth-Moon system and the redshift, but the dark matter field fluid model demonstrates that those two events obey the same law of motion (Eq. 5) and share the same mechanism. Such mechanism should not cause any blurring to the observed light.

### 3. The geometrical aspects of photons

Light is the most mysterious and common phenomenon in the nature, people are always fascinated by the properties of light with all kinds of imagination, such imagination may be more important than the knowledge and can make people thinking in unusual ways without being limited by their knowledge. At 16 years old, Albert Einstein wondered what it would be like to ride on a beam of light. Maxwell described the light as a wave; Einstein described the light as a stream of energy packs which are now called photons, i.e. particles. So the light has wave-particle dual properties. It is quite often that people with different backgrounds and ages wonder what a photon looks like and how big it is. Such question is very interesting, but does not have an answer now, may never have one. However, we can still address some of the geometrical aspects of the photons and interesting information can be extracted out from the above results.

We must accept the facts that photons are real matter objects and each photon is created as a whole entity and travels in the space as a whole entity without falling apart during its journey, therefore there must be some kind of an internal force existed inside the photon to hold "all components" of the photon together and keep it stable during the journey, such internal force is equivalent to the internal force to hold "all components" of an electron together, and is equivalent to the internal force to hold all components of an atom together and is equivalent to the internal force to hold all components of solar system together. In general, when we talk about the particle property, we will naturally think about the size and shape which are associated with particles. For wave-like property, on other hand, the size and the shape may lose their meanings according to quantum mechanics. However, this does not mean that those objects which can be perfected treated by quantum mechanics do not have, at least in time average, the geometrical characteristics such as size and shape. The instant geometries of photons, electrons and other particles may vary rapidly with time, however, in time average, they should have certain stable sizes and shapes, although it is hard to know what such sizes and shapes are. For example, the instant size and shape of hydrogen atom rapidly changes with time due to the fact that the distance and orientation of electron to the proton rapidly changes because of the fast motion of electron around the proton. However, in time average, the electron cloud is distributed around the proton which makes the hydrogen atom have a "ball" shape with the stable average size well represented by the Bohr radius (0.53 Å).

From above results, one can see that the observed redshifts agree with the description of the model very well and photons with all wavelength-band follow the law of motion Eq. 5. As indicated above, the parameter $\alpha$ in Eq. 5 depends on the geometrical characteristics of the object (such as the size, shape, etc) and the $\alpha$ is the same for all photons; therefore, based on such information, we can conclude that all photons have the same geometrical characteristics (size and shape, at least in time average). This means that when a photon travels through the dark matter field fluid, it gradually loses its energy due to the dragging effect of the dark matter field fluid, but its geometry remains the same all the way in its journey. We can further conclude that at least in time average all photons in all inertial reference frames have the same geometry; therefore, photons do not have the length contraction effect. This is a very important property of photons in addition to that all photons have the same speed in all inertial reference frames. In contrast, an object with rest mass has length contraction effect when it is observed in different inertial reference frames with relative motions according to the special relativity.

## 4.     Conclusion

The dark matter field fluid model has been successfully applied to the cosmological redshift, the results deduced from this model agree very well with the observations. The observed cosmological redshift of light depends on both the speed of the emitter and the distance between the emitter and the observer. If the emitter moves away from us, a redshift is observed. If the emitter moves towards us, whether a redshift, a blueshift or no shift is observed will depend on the speed vs. the distance. If the speed is in the range of c(exp[-βD] – 1) < v < 0, a redshift is observed; if the speed equals c(exp[-βD] – 1), no shift is observed; if the speed v < c(exp[-βD] – 1), a blueshift is observed. A redshift will be always observed in all directions for any celestial objects as long as their distance from us is large enough. Therefore, many more redshifts than blueshifts should be observed for galaxies and supernovae, etc in the sky. This conclusion agrees with current observations. The estimated value of the redshift constant β of the dark matter field fluid is in the range of $10^{-3} \sim 10^{-5}$ /Mpc. A large redshift value from a distant celestial object may not necessarily indicate that it has a large receding speed. At least in time average, all photons have the same geometry in any inertial reference frames and do not have length contraction effect.

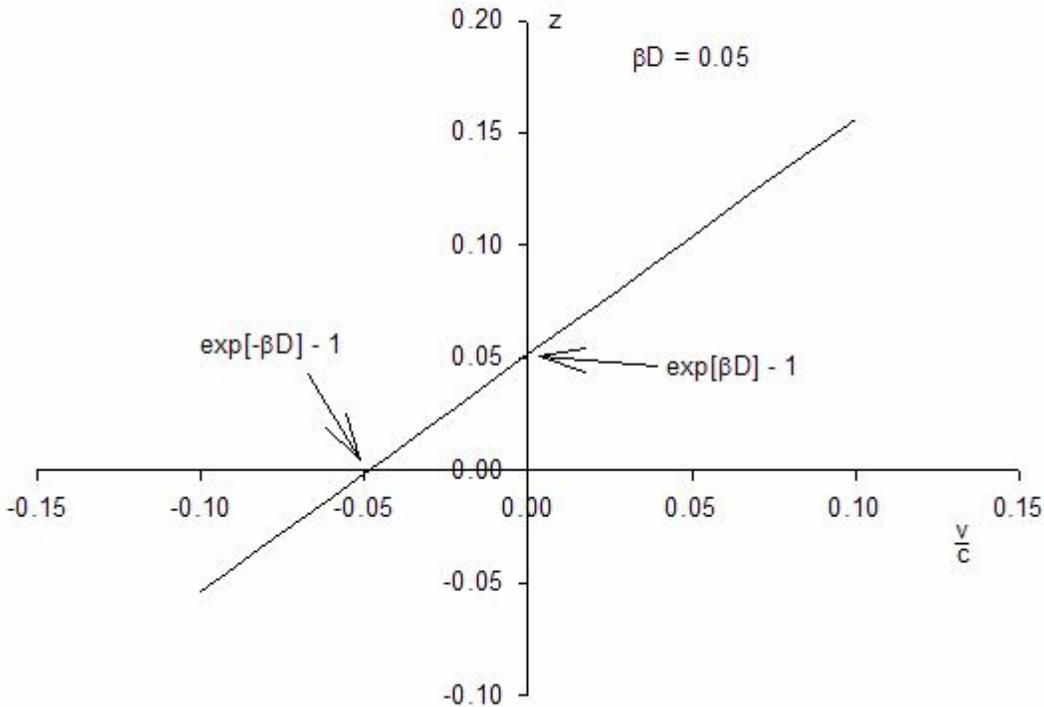

Figure1, the redshift changes with the speed of an emitter

The dependence of observed redshift z on the speed v of the emitter with βD = 0.05. The speed v of the emitter is in unit of speed of light c.